\documentclass[utf8]{frontiersSCNS} 

\usepackage{url,lineno,microtype,subcaption}
\usepackage[onehalfspacing]{setspace}
\usepackage[inline]{trackchanges}

\def\keyFont{\fontsize{8}{11}\helveticabold}
\def\firstAuthorLast{Oliveira {et~al.}} 
\def\Authors{Denny M. Oliveira\,$^{1,2,*}$}

\def\Address{$^{1}$Goddard Planetary Heliophysics Institute, University of Maryland, 
				   Baltimore County, Baltimore, MD, United States
			 $^{2}$Geospace Physics Laboratory, NASA Goddard Space Flight Center, Greenbelt, MD, United States
			 }

\def\corrAuthor{Denny M. Oliveira}
\def\corrEmail{denny@umbc.edu}

\def\dbdt{d$B$/d$t$}
\def\thxn{$\theta_{x_n}$}
\def\Bz{$B_z$}

\usepackage[breaklinks, colorlinks = true,
            linkcolor = SkyBlue,
            urlcolor = SkyBlue,
            citecolor = SkyBlue,
            anchorcolor = white,
            breaklinks]{hyperref}

\begin{document}
\onecolumn
\firstpage{1}

\title[Shock mini review]{Geoeffectiveness of Interplanetary Shocks Controlled by Impact Angles: Past Research, Recent Advancements, and Future Work} 

\author[\firstAuthorLast ]{\Authors} 
\address{} 
\correspondance{} 

\extraAuth{}

\maketitle

\begin{abstract}

	Interplanetary (IP) shocks are disturbances commonly observed in the solar wind. IP shock impacts can cause a myriad of space weather effects in the Earth's magnetopause, inner magnetosphere, ionosphere, thermosphere, and ground magnetic field. The shock impact angle, measured as the angle the shock normal vector performs with the Sun-Earth line, has been shown to be a very important parameter that controls shock geoeffectivess. An extensive review provided by \cite{Oliveira2018a} summarized all the work known at the time with respect to shock impact angles and geomagnetic activity; however, this topic has had some progress since \cite{Oliveira2018a} and the main goal of this mini review is to summarize all achievements to date in the topic to the knowledge of the author. Finally, this mini review also brings a few suggestions and ideas for future research in the area of IP shock impact angle geoeffectiveness.

	\tiny
 	\keyFont{ \section{Interplanetary shocks, shock geometry, geomagnetic activity, geospace response, ionospheric response, ground response}} 
\end{abstract}

\section{Introduction}

	Interplanetary (IP) shocks are solar wind perturbations that directly trigger geomagnetic activity in the magnetosphere-ionosphere-thermosphere (MIT) system. IP shocks rapidly trigger magnetic sudden impulses in geosynchronous orbit \citep{Wang2009}, magnetotail \citep{Huttunen2005}, and on the ground \citep{Smith1986,Echer2005a,Smith2020a,Hajra2020}; enhance field-aligned currents \citep{Belakhovsky2017,Kasran2019}, trigger auroral substorms \citep{Kokubun1977,Zhou2001,Yue2010}, cause dayside auroras \citep{Zhou1999,Tsurutani2001b,Zhou2003}, affect radiation belts \citep{Schiller2016,Bhaskar2021}, excite magnetospheric ultra-low frequency (ULF) waves \citep{Kangas2001,Hartinger2022}; cause geomagnetically induced currents (GICs) \citep{Carter2015,Belakhovsky2017,Tsurutani2021}, ionospheric total electron content \citep{Chen2023}, and thermospheric neutral mass density enhancements that intensify satellite orbital drag \citep{Shi2017,Oliveira2017c,Oliveira2019b}. Therefore, space weather-related effects can be observed in many regions of the MIT system. \par

\begin{figure}
		\centering
		\includegraphics[width = 12cm]{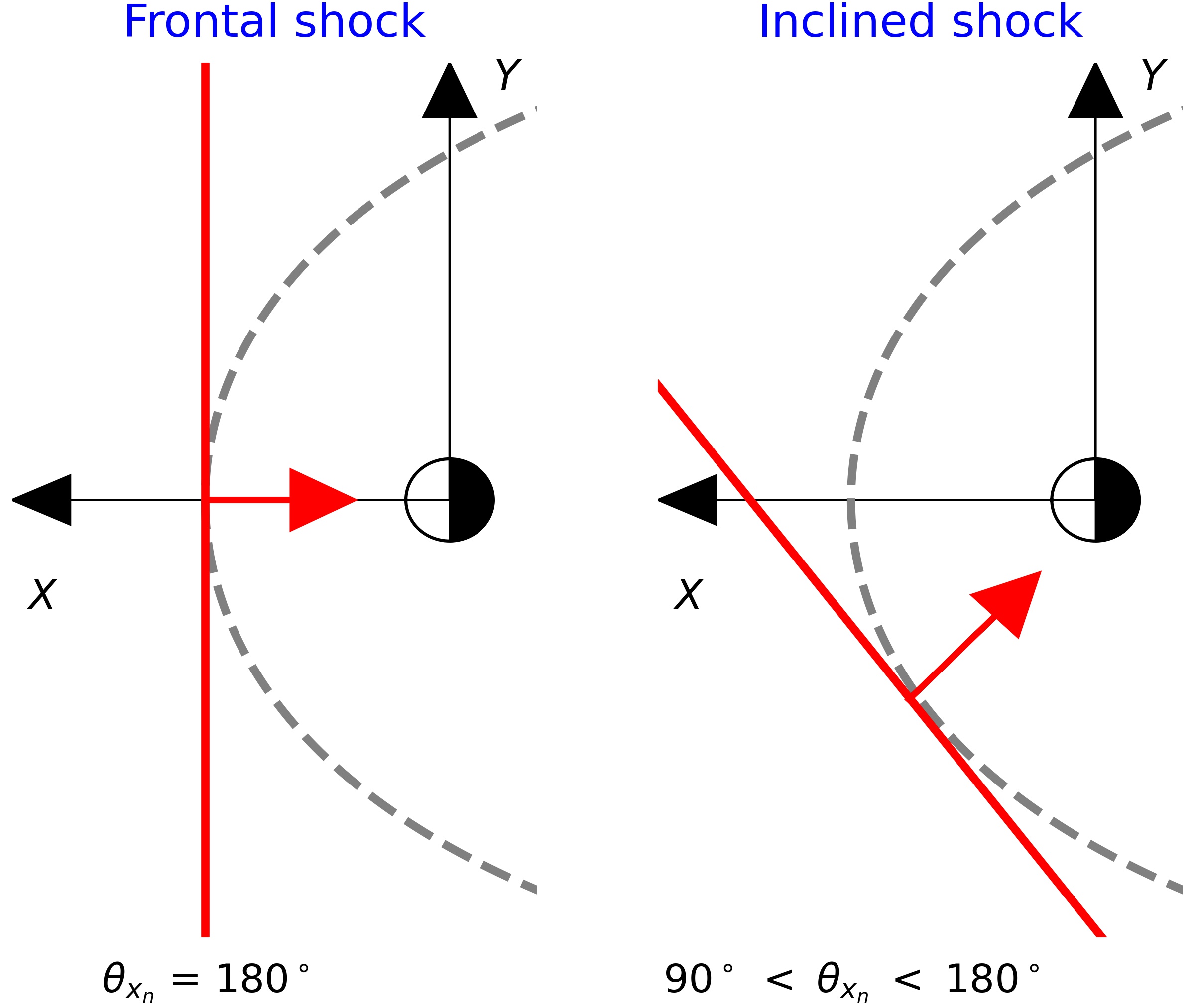}
		\caption{Two schematic representations of shock normal orientations: left, a purely frontal shock (\thxn{} = 180$^\circ$); and right, an inclined shock (90$^\circ$ $<$ \thxn{} $<$ 180$^\circ$). More details of the shock angle, and how it is calculated, can be found in \cite{Oliveira2018a}. The quiet magnetopause (dashed grey lines) was computed by the \cite{Shue1998} empirical model.}
		\label{shocks}
	\end{figure}

	IP fast forward shocks form when the driver gas has a speed relative to the upstream solar wind that is larger than the local magnetosonic wave speed \citep{Landau1960,Priest1981,Kennel1985}. The most common and geoeffective type of shocks is named fast forward shock, which propagates away from the Sun \citep{Tsurutani2011a,Oliveira2017a}. Among several other shock properties, the shock impact angle, \thxn, has been shown to be a very important factor controlling the subsequent geomagnetic activity triggered by shocks \citep[see review by][]{Oliveira2018a}. In this mini review, a head-on shock is indicated by \thxn{} = 180$^\circ$, and an inclined shock has 90$^\circ$ $<$ \thxn{} $<$ 180$^\circ$. As this angle decreases, the shock becomes more inclined with respect to the Sun-Earth line. Figure \ref{shocks} brings a pictorial representation of shock normal orientations. In general, shocks driven by the magnetic cloud (MC) portion of coronal mass ejections (CMEs) are nearly frontal \citep{Klein1982}, while shocks driven by corotating interacting regions (CIRs) are more inclined (with respect to the GSE x axis) since the region of the compression of slow streams by fast streams tend to follow the Parker spiral \citep{Tsurutani2006a}. This angle is computed with the Rankine-Hugoniot conditions which use conservation of energy and momentum across the shock front \citep{Priest1981,Berdichevsky2000}. Since the focus of this mini review is on geomagnetic activity triggered by IP shocks with different inclinations, we recommend the interested reader to consult other references for shock normal computation methods \citep[e.g.,][]{Tsurutani1985b,Schwartz1998,Tsurutani2011a,Oliveira2017a}. \par

	Since the review provided by \cite{Oliveira2018a} on geomagnetic activity triggered by IP shocks with different orientations, some progress has been made in the understanding of this research topic. The main goal of this mini review is to summarize all achievements in the field, including new research, as well as to show that more work is still needed in the area. \par

\section{Magnetic sudden impulses on the ground}

	Perhaps the first clear connection between shock impact angles and the subsequent geomagnetic activity was reported by \cite{Takeuchi2002b}. The authors clearly showed that a very inclined IP shock observed by Wind took an unusually long time to sweep by the magnetosphere while slowly compressing it. The slow and gradual increase in total pressure (thermal plus magnetic pressures) observed by Geotail in the magnetosphere and the slow and gradual magnetic field increase observed by magnetometers on the ground were associated with the large shock inclination of the IP shock. \cite{Takeuchi2002b} argued that these effects were caused by the asymmetric compression of the magnetosphere which led currents in the magnetosphere-ionosphere system to respond asymmetrically and slowly. Additionally, \cite{Takeuchi2002b} introduced the concept of a geoeffective magnetosphere distance, which is the distance swept by the shock along its normal while the magnetosphere is being compressed. These results were later confirmed with numerical magnetohydrodynamic  simulations by \cite{Guo2005}, who simulated two similarly strong shocks with different inclinations, one being frontal, and the other inclined. The authors showed that magnetospheric magnetic field and ionospheric cross polar cap potential reached similar final values after the compression by the shocks, but the effects caused by the inclined shock took longer to take place. \cite{Wang2005} conducted numerical simulations of shocks with similar normal orientations and noted similar response (slow for the inclined shock, and rapid for the frontal shock) of the magnetic field at geosynchronous orbit. \par

	These previous observation and numerical simulation studies were later confirmed by \cite{Wang2006a}, who performed a statistical analysis of ground magnetometer response to shocks with different inclinations. Their main conclusion was that the faster and the more frontal the shock, the shorter the sudden impulse rise time. \cite{Selvakumaran2017} performed a statistical analysis study of the same topic and reported on similar conclusions. \cite{Samsonov2015} noted dawn-dusk asymmetries in a numerical simulation of an inclined shock impact on the magnetosphere. The authors noted an overshoot in the field calculated by an artificial ground magnetometer located in the dusk sector where the shock first contacted the magnetopause, but an overshoot in another station in the dawn sector did not clearly occur. The lack of dawnside overshoots in the simulation and observations was then explained by the fact that the dawnside compression was preceded and followed by magnetopause expansions. These results were also confirmed with magnetic field observations by real ground magnetometers located in the same sectors as the artificial stations. \cite{Samsonov2015} explained this dawn-dusk asymmetry with the inclined shock having a very strong solar wind velocity $v_y$ component. As theoretically noted by \cite{Samsonov2011a}, a very strong solar wind $v_y$ component greatly increases the solar wind ram pressure in the y direction, leading to the strong dawn-dusk asymmetry noted in ground magnetic field observed and simulated by \cite{Samsonov2015}. Historical observations of very intense sudden impulses on the ground suggest that such response was caused by impacts of very fast and nearly frontal shocks on the Earth's magnetosphere \citep{Araki2014,Oliveira2017d,Love2019a,Love2019b,Hayakawa2020a,Hayakawa2020b,Hayakawa2021a,Hayakawa2021c}. \par

	The geoeffective magnetosphere distance concept was investigated by \cite{Rudd2019}, who performed a statistical study with more than 500 IP shocks with different inclinations. The authors reported that shorter sudden impulse rise times were associated with faster and more frontal shocks, confirming the experimental and simulation results abovementioned \citep{Wang2005,Wang2006a}. \cite{Rudd2019} used each respective shock speed and the subsequent sudden impulse rise time to find that faster nearly frontal shocks traveled relatively smaller distances while compressing the magnetosphere in comparison to slower highly inclined shocks. These results confirmed the prediction of \cite{Takeuchi2002b}.

\section{Substorm occurrence and intensities}

	\cite{Oliveira2014b} conducted global numerical simulations of the impact of three IP shocks with different inclinations and strengths on the Earth's magnetosphere. Two shocks were inclined (\thxn = 150$^\circ$), and the other shock was purely frontal (\thxn = 180$^\circ$). The head-on shock and one of the inclined shocks were moderate events, and the other inclined shock was a strong event. The shock normal inclination was with respect to the GSE meridional (xz) plane. The authors found that the overall geomagnetic response was faster and more intense in the head-on shock case. Results showed that a clear substorm activity according to \cite{Akasofu1964a}'s substorm development framework was seen a few minutes after shock impact only in the case of the frontal shock. The authors concluded that such response occurred due to the rapid and effective compression of the magnetotail that effectively unloaded electromagnetic energy previously stored in the magnetotail, which did not clearly occur in the case of the inclined shocks, even in response to the strong inclined shock. \par

	Later on, \cite{Oliveira2015a} performed a statistical study of ground magnetometer response to more than 400 shocks with different inclinations observed by the Wind and Advanced Composition Explorer satellites at L1. The authors looked particularly at the SuperMAG westward auroral electrojet index SML for each event. \cite{Oliveira2015a} found that the faster and the more frontal the shock, the more intense the SML response, which indicates that the shock impact angle controls the subsequent substorm activity. Similar results for nightside auroral power intensity were reported by \cite{Oliveira2016a}. Therefore, these experimental results confirmed the predictions reported by \cite{Oliveira2014b} with respect to effects of IP shock inclinations on the subsequent substorm activity. \par

	It is important to note that effects of magnetospheric preconditioning caused by southward IMF \Bz{} are an important factor for substorm triggering \citep{Kokubun1977,Zhou2001,Yue2010,Hajra2018a,Sitnov2019}. As shown by \cite{Oliveira2014b} with simulations and \cite{Oliveira2021b} with observations, nearly head-on shocks enhance the upstream IMF \Bz{} component causing the downstream IMF \Bz{} component to become even more negative which in turn intensifies the energetics of substorms, as first shown by \cite{Tsurutani1988}. This sudden and explosive tail energy release can explain why supersubstorms are way more intense and last longer than intense substorms \citep{Tsurutani2023}. Therefore, the possible role of shock impact angles in supersubstorm triggering is a topic that is yet to be investigated.

\section{Ultra-low frequency waves}

	In their numerical simulations, \cite{Oliveira2014b} also reported on the effects of IP shock impact angles on the occurrence of ULF waves. They noted a clear ULF wave perturbation with period $\sim$5 minutes triggered by the frontal shock, but such response was not clearly seen in either of the inclined shocks, even in the case of the strong inclined shock impact. \cite{Oliveira2014b} attributed this effect to an effective excitation of cavity modes \citep{Samson1992,Hughes1994,Lakhina2020} in the magnetosphere-ionosphere system by the symmetric compression caused by the frontal shock. The inclined shocks, on the other hand, were not able to trigger these cavity modes due to the asymmetric magnetospheric compressions. These simulation results were used by \cite{Belakhovsky2017} to explain the lack of ULF wave activity after the impact of a highly inclined shock on the magnetosphere. \cite{Takahashi2018a} also noted intense ULF wave activity following the impact of an almost head-on shock on the magnetosphere. \cite{Baker2019} reported that nearly frontal shocks triggered Pc5 waves with more intense wave power spectra in comparison to highly inclined shocks. \par

	A comparative study of ULF activity triggered by two IP shocks with similar strengths, but with very distinct inclinations, was performed by \cite{Oliveira2020d}. The authors looked at ULF waves in the geospace and on the ground and observed stronger wave response in the case of the nearly head-on shock impact. Such response was characterized by wave amplitude and power spectra. In addition, \cite{Oliveira2020d} suggested that a frontal shock can only excite odd-mode waves in the magnetosphere, whereas an inclined shock can excite both odd- and even-mode waves in the magnetosphere. This is supported by peaks in power spectral density that occurred periodically and ``out of phase" when the results of the two shocks are compared. These results confirmed the predictions of \cite{Oliveira2014b}, but a comprehensive statistical study of ULF wave response to IP shocks with different orientations in the geospace and on the ground is yet to be conducted to confirm the statistical properties of such wave response.

\section{Ground \dbdt{} variations}

	Ground \dbdt{} variations are the space weather drivers of GICs that can be detrimental to power infrastructure and transmission lines \citep{Pirjola2002,Oliveira2017d,Ngwira2019}. The first connection between shock impact angles and the subsequent ground \dbdt{} variations was established by \cite{Oliveira2018b}. The authors found that the faster and the more frontal the shock, the more intense the resulting ground \dbdt{} variation within 20 minutes after shock impact, particularly in high latitude regions. With respect to low-latitude response, \cite{Oliveira2018b} reported that high \dbdt{} variations surpassing the threshold of 100 nT/min at the equator was associated with nearly head-on shocks that struck the magnetosphere around noon local time. Such \dbdt{} intensifications resulted from enhancements of the dayside ionospheric equatorial electrojet current, as previously suggested by \cite{Carter2015}. \par

	The \dbdt{} response during two very intense substorms triggered by two IP shocks with different inclinations was investigated in a comparative case study by \cite{Oliveira2021b}. \cite{Hajra2018a} defined a supersubstorm as an event with minimum SML $<$ --2500 nT, and an intense substorm as an event with minimum SML in the interval --2500 nT $<$ SML $\leq$ --2000 nT. \cite{Oliveira2021b} observed that a nearly frontal shock induced a supersubstorm, whereas a highly inclined shock induced an intense substorm, both occurring after periods with southward IMF \Bz{} conditions \citep{Hajra2018a}. The authors also observed that the \dbdt{} response was stronger and faster in the case of the nearly frontal shock, whereas the \dbdt{} response was weaker and occurred later in the case of the highly inclined shock. Additionally, geographic areas with \dbdt{} peaks surpassing 1.5 nT/s and 5 nT/s, were larger and peaked earlier during the supersubstorm. Such \dbdt{} thresholds can be detrimental to power transmission lines and infrastructure \citep{Molinski2000,Pulkkinen2013}. \cite{Oliveira2021b} showed that these \dbdt{} peaks correlated in time with intense energetic particle injections observed by Time History of Events and Macroscale Interactions during Substorms (THEMIS) and Los Alamos National Laboratory spacecraft in the tail around the magnetic midnight. The energetic particle injections triggered by the nearly head-on shock occurred faster and were more intense, but the energetic particle injections triggered by the highly inclined shock were less intense and occurred slower in two phases represented by a double peak caused by the asymmetric magnetospheric compressions. A clear intense auroral brightening and poleward expansion of the auroral oval were observed with THEMIS all-sky images in the case of the supersubstorm, but these effects were not clearly observed in the case of the intense substorm. \par

	Inter-hemispheric asymmetries in the ground \dbdt{} response triggered by shocks with different inclinations were reported by \cite{Xu2020a}. IP shock impact angles were expressed in solar magnetic coordinates to highlight north-south magnetic field response \citep{Laundal2017a}. \cite{Xu2020a} used conjugated magnetic field data provided by ground magnetometers in Antarctica and Greenland to find that, in general, the hemisphere that is first struck by the shock shows the first and more intense \dbdt{} response.

\section{Bow shock, field-aligned currents, and cross-tail currents}
	
	The Earth's bow shock response to the impact of an inclined IP shock was theoretically investigated by \cite{Grib2006}. They considered the impact of a shock with \thxn = 135$^\circ$ on the duskside bow shock. By numerically solving the Rankine-Hugoniot equations at different points along the bow shock surface in the dusk sector, \cite{Grib2006} found that discontinuity structures propagated along the magnetosheath with strong dawn-dusk asymmetries after the shock impact, with the local density being nearly $\sim20$\% higher in the dusk flank with respect to the dawn flank. In a similar analysis, \cite{Grib2016} found plasma irregularities in front of the bow shock right before the impact of inclined solar wind discontinuities on the bow shock. \par

	In their numerical simulations, \cite{Oliveira2014b} found faster and more intense field-aligned current (FAC) response in the case of the frontal shock in comparison to the other two inclined shocks. The authors noted intense FAC activity around the magnetic midnight in response to the frontal shock. \cite{Oliveira2014b} associated this FAC response to the substorm activity noted by the authors in the simulations, and such effects did not occur in response to the inclined shocks. \cite{Selvakumaran2017} found similar results in their numerical simulations of shocks with different orientations. \par

	\cite{Shi2019b} used FAC data provided by the Active Magnetosphere and Planetary Electrodynamics Response Experiment program to investigate how the shock impact angle affects the subsequent current dynamics. The authors found that FACs showed rapid intensifications after the shock impact, and FACs developed faster and stronger for the more frontal shocks. \cite{Shi2019b} also found that nearly frontal shocks caused stronger currents particularly in the dayside sector near local noon, dawnside Region I currents, and duskside Region II currents in comparison to highly inclined shocks. \par

	In the numerical simulation conducted by \cite{Samsonov2015}, it was found that the cross-tail current was highly deformed after the impact of an inclined IP shock on the magnetosphere. This effect was caused by the high $v_y$ component of the solar wind downstream region that led to an elevated dynamic pressure enhancement in the east-west direction \citep{Samsonov2011a}. \cite{Grygorov2014} used multi-satellite data to observe the effects caused by an IP shock on the magnetotail ($X$ = $\sim-$240$R_E$). The authors found that the tail was highly deflected ($\sim30^\circ$) as a result of the magnetosphere compression by an inclined shock (\thxn $\sim150^\circ$) which drove a very intense solar wind ram pressure in the east-west direction \citep{Samsonov2011a}.

\section{Conclusions and future work}

	In this mini review, with some updates to the review provided by \cite{Oliveira2018a}, I have summarized many experimental and modeling studies concerning geomagnetic activity triggered by IP shocks with different impact angles. To the author's knowledge, this review is the most up-to-date report on this topic. As expected, most studies reported here concluded that the faster and the more frontal the shock, the more intense the subsequent geomagnetic activity. Moreover, the solar wind flow angle also plays a significant role in the subsequent geoeffectiveness even during periods of non-shocked solar wind. For example, \cite{Rout2017} found that periodic equatorial ionospheric response to CIRs are higher when the solar wind azimuth flow angle is smaller than $\sim$6$^\circ$ at L1. Additionally, \cite{Cameron2019a} found higher geomagnetic activity indicated by ground magnetometer measurements was higher when solar wind front normals lied in the equatorial plane aligned with the Parker spiral at an angle of 45$^\circ$. Although this mini review shows several advancements obtained by many efforts, I still identify a few areas of potential investigations of shock impact angle control of the subsequent geoeffectiveness. They are: 

	\begin{enumerate}

		\item {\it Radiation belt response}. How do IP shock impact angles control particle acceleration in the inner magnetosphere? And particle population flows in the inner magnetosphere? How does the shock impact angle control enhancements and dropouts of, e.g., electron densities in the inner magnetosphere? How does the shock impact angle control losses of magnetospheric relativistic electron fluxes \citep[e.g., ][]{Tsurutani2016,Hajra2018b}?

		\item {\it ULF wave activity.} Do shock impact angles control ULF wave (Pc4-5, $\sim$2-22 mHz) amplitude/power spectra at different locations, and how does Pc4-5 wave energy spread throughout the magnetosphere as a function of local time for different impact angles? Do shock impact angles control the field-aligned mode structure of ULF waves, as suggested by \cite{Oliveira2020d}? 

		\item {\it Statistical properties of ground \dbdt{} variations}. A logical step following the comparative study of \cite{Oliveira2021b} is to perform statistical analyses of ground \dbdt{} variations during substorms induced by shocks with different inclinations. What are the conditions in geospace that trigger substorms under asymmetric magnetospheric compressions? What is the latitudinal extent and intensity of \dbdt{} variations as a function of shock impact angles?

		\item {\it Thermospheric neutral mass density response}. How does the shock impact angle affect the high-latitude thermospheric neutral mass density response in low-Earth orbit? How does this density impact the subsequent satellite orbital drag? How does the impact angles of CIR-driven shocks impact the immediate density response and the subsequent azimuth solar wind flow and the resulting satellite orbital drag effects?

	\end{enumerate}

Finally, I would like to suggest modelers to consider IP shock inclinations in their numerical simulations. As argued by \cite{Welling2021}, ground \dbdt{} variations were intensified and reached very low magnetic latitudes after the impact of a very fast and head-on CME on the magnetosphere. Moreover, this is also supported by previous observations showing that most shocks observed in the solar wind are moderately inclined shocks, with shock normal deviations of $\sim$40$^\circ$ with respect to the Sun-Earth line \citep{Oh2007,Kilpua2015a,Oliveira2018a,Rudd2019}.

\section*{Conflict of Interest Statement}

	The author declares that the research was conducted in the absence of any commercial or financial relationships that could be construed as a potential conflict of interest.

\section*{Author Contributions}

	This mini review article was written by the author without any direct contributions from others.

\section*{Funding}
	
	This work was possible thanks to the financial support provided by the NASA HGIO program through grant 80NSSC22K0756.

\bibliographystyle{frontiersinSCNS_ENG_HUMS} 
\bibliography{Oliveira_main}

\end{document}